\documentclass[prl,showpacs,twocolumn]{revtex4}%
\input{psfig.sty}
\usepackage{amsmath}
\usepackage{graphicx}
\usepackage{amsfonts}
\usepackage{amssymb}%

\begin{document}
\title{Entanglement of remote atomic qubits}
\date{\today }
\author{D. N. Matsukevich, T. Chaneli\`{e}re, S. D. Jenkins, S.-Y. Lan, T.A.B. Kennedy, and A. Kuzmich}
\affiliation{School of Physics, Georgia Institute of Technology, Atlanta, Georgia 30332-0430}
\begin{abstract}
We report observations of entanglement of two remote atomic qubits, achieved by generating an entangled state of an atomic qubit and a single
photon at Site {\it A}, transmitting the photon to Site {\it B} in an adjacent laboratory through an optical fiber, and converting the photon
into an atomic qubit. Entanglement of the two remote atomic qubits is inferred by performing, locally, quantum state transfer of each of the
atomic qubits onto a photonic qubit  and subsequent measurement of polarization correlations in violation of the Bell inequality $|S| \leq 2$. We
experimentally determine $S_{exp} =2.16 \pm 0.03$. Entanglement of two remote atomic qubits, each qubit consisting of two independent spin wave
excitations, and reversible, coherent transfer of entanglement between matter and light, represent important advances in quantum information
science.
\end{abstract}
\maketitle

Realization of massive qubits, and their entanglement, is central to practical quantum information systems \cite{briegel,duan,ekert}. Remote
entanglement of photons can now be achieved in a robust manner using the well-developed technology of spontaneous parametric down-conversion
\cite{zeilinger,boschi,aspelmeyer}, with propagation to remote locations by means of optical fibers. Photons, however, are difficult to store for
any appreciable period of time, whereas qubits based on ground-state atoms have long lifetimes. Local entanglement of massive qubits has been
observed between adjacent trapped ions \cite{wineland} and between pairs of Rydberg atoms in a collimated beam \cite{haroche}. In order to
entangle qubits at remote locations the use of photons as an intermediary seems essential \cite{cabrillo,bose,sorensen,duan1}. Photons also offer
some flexibility as information carriers as they can propagate in optical fiber with low losses. The creation, transport, storage, and retrieval
of single photons between remote atomic ensembles located in two different laboratories was recently reported \cite{chaneliere}. The first step
in creating remote entanglement between massive qubits is to entangle one such qubit with the mediating light field, which is then directed
towards the second qubit via an optical fiber. There have recently been important advances towards this goal by demonstrating entanglement of a
photon with a trapped ion \cite{blinov}, with a collective atomic qubit \cite{matsukevich,matsukevich1}, and with a single trapped atom
\cite{weber}.

A promising route towards the creation and application of long-lived qubit entanglement in scalable quantum networks was proposed by Duan, Lukin,
Cirac, and Zoller  \cite{duan,duan1}. These atomic qubits rely on collective atomic states containing exactly one spin excitation. For useful
quantum information processing two orthogonal spin wave excitation states $\hat s_+^{\dagger }|0\rangle _a, \hat s_-^{\dagger }|0\rangle _a$ are
needed for the logical states of a qubit \cite{duan}, where $|0\rangle _a$ represents the collective atomic ground state.  We note that the two
states $|0\rangle _a$ and $ \hat s^{\dagger }|0\rangle _a$, do not appear to constitute a practically useful qubit \cite{wiseman}. In the
experiment of Ref.\cite{chaneliere} each of the two remote ensembles only contained one logical state, since the atomic ground state component
does not serve this purpose. Entanglement of continuous atomic variables in two separate atomic ensembles has been reported \cite{polzik}, as
appropriate for continuous-variable quantum information processing, but not for qubit entanglement.

In two recent experiments, collective atomic qubits were generated using cold atomic ensembles \cite{matsukevich,matsukevich1}. In the first of
these the logical states were single spin wave excitations (ideally, $\hat s_+^{\dagger }|0\rangle _a, \hat s_-^{\dagger }|0\rangle _a$), in
either one of two distinct atomic ensembles inside a high vacuum chamber. In the second experiment, two orthogonal spin waves of a single cold
ensemble represented the logical qubit states \cite{matsukevich1}.  The experiments \cite{matsukevich,matsukevich1} realized a single atomic
qubit system, but did not address the issue of entanglement of atomic qubits.

While remote entanglement of atomic qubits has not been previously demonstrated, Refs. \cite{matsukevich,matsukevich1} realized two basic
primitives of a quantum network: (a) entanglement of photonic and atomic qubits, and (b) quantum state transfer from an atomic to a photonic
qubit. The crucial additional ingredient is the reverse operation, the conversion of a photonic qubit into an atomic qubit. This enables the
transfer of atom-photon entanglement into remote atomic qubit entanglement.

Here we report remote atomic qubit entanglement using cold atomic clouds of $^{85}$Rb confined at Sites {\it A} and {\it B}, as shown in Fig.~1.
These sites are situated in separate laboratories and linked by an optical fiber. A notable distinction between the two nodes is that the qubit
generated at Site {\it A} is written on an unpolarized atomic ensemble, as in Ref. \cite{matsukevich1}, whereas at Site {\it B} the atomic
ensemble is prepared, ideally, in the ($m=0$) Zeeman state of the $F=2$ ground level by optical pumping. All the light fields responsible for
trapping and cooling of the atoms, as well as the quadrupole magnetic fields at both sites, are shut off during the period of the protocol. The
ambient magnetic field at each site is compensated by three pairs of Helmholtz coils, and a bias field of $0.2$G is added at Site {\it B} for the
purpose of optical pumping.

Our protocol starts with the generation of an entangled state of a signal photon and a collective atomic qubit  at Site {\it A}, achieved through
Raman scattering of a classical laser write pulse. The state can be represented schematically as
\begin{eqnarray}
|\Psi\rangle &=& |0\rangle _a|0\rangle _f + \chi (\cos \eta |+\rangle _a  |+\rangle_f   + \sin \eta|-\rangle _a |-\rangle _f) \nonumber \\
&\equiv& |0\rangle _a|0\rangle _f + \chi |\psi\rangle,
\end{eqnarray}
where $|+\rangle _f \equiv \hat a_+^{\dagger }|0\rangle _f$ and $|-\rangle _f \equiv \hat a_-^{\dagger }|0\rangle _f$ are the normalized states
of positive and negative helicity of the signal photon, $|0\rangle _f$ is the field vacuum state, $|\pm \rangle _a \equiv \hat s_{\pm }^{\dagger
}|0\rangle _a$ describes the two logical qubit states, corresponding to non-symmetric collective atomic modes \cite{kuzmich3}, and $\chi << 1$.
The asymmetry angle $\eta =0.81 \pi/4$ \cite{matsukevich1}.  Eq.(1) represents probabilistic entanglement generation, where ideally for each
signal photon emission event, an entangled atomic qubit is created in the atomic ensemble \cite{blinov,duan}. Since we deal with an unpolarized
atomic ensemble, the state of the system is more rigourously described by a density operator as discussed in Ref.\cite{matsukevich1}.

\begin{figure}[htp]
\begin{center}
 \leavevmode  \psfig{file=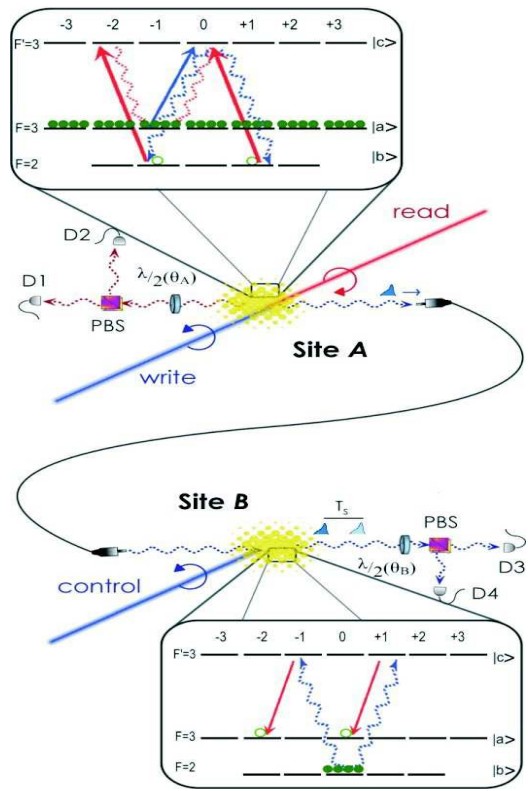,height=4.1in,width=2.7in}
\end{center}
\caption{ A schematic diagram of our experimental setup. Two cold atomic ensembles of $^{85}$Rb, an unpolarized sample at Site {\it A}, and a
spin-polarized sample at Site {\it B}, separated by 5.5 m, are connected by a single-mode fiber. The insets show the structure and the initial
populations of the atomic levels for the two ensembles. An entangled state of a collective atomic qubit and a signal field is generated at Site
{\it A} by Raman scattering of the write laser field. The orthogonal helicity states of the generated signal field are transmitted via optical
fiber from Site {\it A} to Site {\it B}, where they are converted to orthogonal collective atomic excitations, stored for a duration $T_s$, and
subsequently converted into an idler field by adiabatic variation of the control field amplitude. The atomic qubit at Site {\it A} is similarly
converted into an idler field by a read laser pulse, counterpropagating with respect to the write pulse. For polarization analysis, each idler
field propagates through a quarter-wave plate (not shown), a half-wave plate ($\lambda /2$) and a polarizing beamsplitter (PBS). Polarization
correlations of the idler fields are recorded by photoelectric detection using the single photon detectors D1-D4.}\label{TQ}
\end{figure}

The orthogonal polarization modes of the signal field produced at Site {\it A} are directed along the optical fiber to Site {\it B}. As the
signal field propagates from Site {\it A} to Site {\it B}, it passes through two quarter wave plates, causing the transformation of the signal
field operators $  \hat{a}_{\pm} \rightarrow \pm \hat{a}_{\mp}$. The signal field propagation in the atomic medium at Site {\it B} is controlled
by an additional laser field (control) through the process of electromagnetically-induced transparency (EIT)
\cite{harris,scully,hau0,fleischhauer,phillips,hau}.

We implement the storage phase at Site {\it B}, by adiabatically reducing the control field amplitude to zero, while the signal pulse lies within
the cloud. The orthogonal atomic spin wave excitations thereby created in the spin-polarized gas constitute the logical states of the atomic
qubit. In order to convert the signal field qubit into a collective atomic qubit, it is necessary that the optically thick atomic sample supports
EIT for both field helicities \cite{fleischhauer}. To this end, we optically pump the atomic cloud at Site {\it B} using a linearly polarized
field resonant to the $F=2 \leftrightarrow F^{\prime} =2$ transition of the $D_1$-line, and an additional repumping field resonant to the $F=3
\leftrightarrow F^{\prime} =3$ transition of the $D_2$-line. We measured the optical thickness $d \simeq 8$ for both circular components of the
signal field.

By switching off the control field over a period of about 20 ns, the photonic qubit is converted into an atomic qubit. At this stage remote
atomic qubits should have been created at Sites {\it A} and {\it B}. Atoms at Site {\it B} should, ideally, be prepared in a single Zeeman $m=0$
state of the $F=2$ hyperfine ground level (lower inset in Fig.~1). In practice the pumping is not perfect, possibly due to radiation trapping in
the optically thick atomic medium \cite{tupa}. We measure lower storage and retrieval efficiency for the negative helicity signal component
compared with that of the positive helicity component (3\% vs 8\%). Numerical simulations indicate that the discrepancy between the efficiencies
is consistent with a residual population in the $|F=2,m=-2\rangle$ atomic state at the $10 \%$ level \cite{jenkins1}.  This results in
undesirable absorption of the signal field with negative helicity.

 The signal photon of helicity $\alpha=\pm1$ is stored in the ensemble at Site
{\it B} with   efficiency $\epsilon_\alpha$. After a storage time $T_s$, the non-vacuum component of the state of the two ensembles is given by
the following density operator: $\hat{\rho} = (1-\epsilon)\hat{\rho}_A + \epsilon \hat{\rho}_{AB}$, where the component $\hat{\rho}_A$ describes
the state of single excitation at Site {\it A}, and is expressed by
  \begin{equation}
    \hat{\rho}_A = \frac{1-\epsilon_{-}}{1-\epsilon}\cos^2\eta
                   \hat{s}_{A+}^{\dag} \hat{\rho}_{vac} \hat{s}_{A+}
         + \frac{1-\epsilon_{+}}{1-\epsilon}\sin^2\eta
           \hat{s}_{A-}^{\dag} \hat{\rho}_{vac} \hat{s}_{A-},
  \end{equation}
where $\hat{\rho}_{vac}$ is the product of the ground state atomic density operators for the ensembles at Sites {\it A} and {\it B}. The density
operator $\hat{\rho}_{AB} = \hat{\Psi}^{\dag}_{AB}(T_s) \hat{\rho}_{vac} \hat{\Psi}_{AB} (T_s)$  in the two-qubit sub-space represents an
entangled atomic state where
  \begin{equation}
    \hat{\Psi}_{AB}^{\dag}(T_s) = e^{i\phi(T_s)}\cos\eta'
                           \hat{s}^{\dag}_{A+}\hat{s}^{\dag}_{B-}
               -\sin\eta' \hat{s}^{\dag}_{A-}\hat{s}^{\dag}_{B+}
  \end{equation}
with $\cos\eta' = \sqrt{\epsilon_{-} / \epsilon} \cos\eta$,  and $\epsilon = \epsilon_{-}\cos^2\eta + \epsilon_{+}\sin^2\eta$ is the average
efficiency of photon storage at Site {\it B}.  The phase $\phi(t) = -2(g \mu_B/\hbar)B_0t$ is induced by the applied magnetic field $B_0=0.2$G
oriented along the propagation axis at Site {\it B}, where $g$ is the Land\'{e} g-factor for hyperfine level with $F=3$.

 Ideally, entanglement should have been created between the collective atomic qubits at Sites {\it A} and {\it B}. After a storage time $T_s$, the
remote collective atomic excitations are converted by quantum state transfer into idler fields emanating from Sites {\it A} and {\it B}, using a
read laser pulse at Site {\it A} and by reactivating the control field at Site {\it B} \cite{matsukevich,matsukevich1}. The resulting idler-idler
photoelectric correlations may be calculated using the effective two-photon state
  \begin{equation}
    |\Psi_2\rangle = \cos\eta_f |HV\rangle + e^{i\phi_f}\sin\eta_f |VH\rangle
  \end{equation}
where $|HV \rangle = \hat{a}_{A,H}^{\dag}\hat{a}_{B,V}^{\dag}|0\rangle _f$ and $|VH\rangle = \hat{a}_{A,V}^{\dag}\hat{a}_{B,H}^{\dag}|0\rangle
_f$, and the subscripts $A$ and $B$ indicate the idler mode at the respective site. We omit higher-order terms in photon number
\cite{chaneliere}.

The phase $\phi _f$, which includes the contributions due to the Larmor precession $\phi(T_s)$, the light phase shifts in the atomic media, and
various optical elements, is introduced as an adjustable parameter. The mixing angle $\eta_f$ is determined by the relative efficiencies with
which the two qubit states are transferred from the atomic ensembles to the idler fields.  If we assume equal transfer efficiencies at Site {\it
A}, we find $\cos\eta_f = \sqrt{\epsilon_{B-}/\epsilon_B}\cos \eta$, where $\epsilon_B = \epsilon_{B-}\cos^2 \eta + \epsilon_{B+} \sin^2 \eta$
and $\epsilon_{B\pm}$ is the combined storage and retrieval efficiency for a photon of helicity $\pm$ at Site {\it B}. Measurements of these
efficiencies give $\epsilon_{B+} = 0.08$, and $\epsilon_{B-} = 0.03$. With $\eta = 0.81\pi/4$ fixed by the atom-photon entanglement process at
Site {\it A} \cite{matsukevich1} we get $\eta_f = 1.12 \pi/4$. Our experimental data, including those displayed in Fig.~3, are consistent with
this value of $\eta _f $ and $\phi _f \ll 1$.

The above arguments are clearly conditional on the generation of the signal qubit. According to Eq.(1), the corresponding probability scales as
$\chi ^2$, and this determines the efficiency of the probabilistic entanglement generation. However, as Duan {\it et al.} point out \cite{duan},
quantum network protocols eliminate the vacuum component of Eq.(1) and only the entanglement characteristics of $|\psi\rangle$ are relevant
\cite{kimblenk,response}. In our experiment, atomic qubits were stored for a time $500$ ns at Site {\it A} and $200$ ns at Site {\it B}. It
should be possible to extend the qubit storage times to longer than 10 $\mu$s, as the single-quanta storage results suggest \cite{chaneliere}.

The measurement of the atomic qubits is performed by quantum state transfer onto the idler fields at both sites, at both sites, using the read
laser pulse at Site {\it A} and the control laser pulse at Site {\it B}. The polarization state of either idler field is measured using a
polarizing beamsplitter and two single photon detectors, D1, D2 for Site {\it A} and D3, D4 for Site {\it B} (additional technical details are
given in Refs.\cite{matsukevich,matsukevich1,chaneliere}). Polarization correlations between the idler fields produced at the remote sites are
recorded and analyzed for the presence of entanglement. The contributions of the vacuum and single photon idler excitations are excluded in the
observed photoelectric coincidences between the remote sites \cite{kimblenk,response}. Since quantum state transfer is a local process, it cannot
generate entanglement. Hence, observation of idler field entanglement confirms probabilistic entanglement of the two remote atomic qubits. We
denote the number of such coincidences between detector $Dn $, $n =1,2$ at Site {\it A} and detector $Dm$, $m =3,4$ at Site {\it B} by  $
C_{nm}\left(\theta _A,\theta _B\right) $. Here $\theta_A$ and $\theta_B$ are the angles by which polarization is rotated by the half-waveplates
at these Sites.

\begin{figure}[htp]
\begin{center}
\leavevmode  \psfig{file=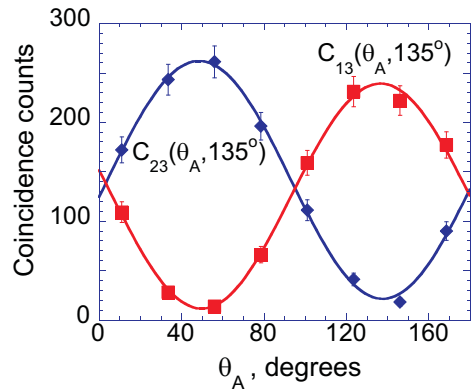,height=2.0in,width=2.4in}
\end{center}
\caption{Measured coincidence fringes  $C_{n3}(\theta _A, \theta _B)$ as a function of $\theta _A$, for $\theta _B =135^{\circ }$, $n=1$,
diamonds, $n=2$, squares. The curves are sinusoidal fits to the data. Each point is acquired for 15 minutes. The effective repetition rate is 108
kHz, each trial takes 1.1 $\mu$s. }\label{Fig2}
\end{figure}

The two-particle interference produces a high-visibility sinusoidal fringe pattern for the coincidence rates $ C_{nm}\left(\theta _A,\theta
_B\right) $, which is characteristic of entangled particles. Fig.~2 shows measured coincidence fringes for some representative angles. We
calculate the coincidence rates $ C_{nm}\left(\theta _A,\theta _B\right) $ to be
 \begin{eqnarray}
    C_{13}(\theta_A,\theta_B) &\propto& \epsilon_1\epsilon_3|(\cos\eta_f + e^{i\phi_f} \sin\eta_f)
                                  \sin(\theta_B+\theta_A)\nonumber \\
                  &+&(\cos\eta_f - e^{i\phi_f} \sin\eta_f)
                                  \sin(\theta_B-\theta_A) |^2,
  \end{eqnarray}
where $\epsilon _m$ is the overall efficiency (including propagation losses) for detector $D_m$, and similar expressions for the other three
rates \cite{jenkins1}.

\begin{figure}[htp]
\begin{center}
\leavevmode  \psfig{file=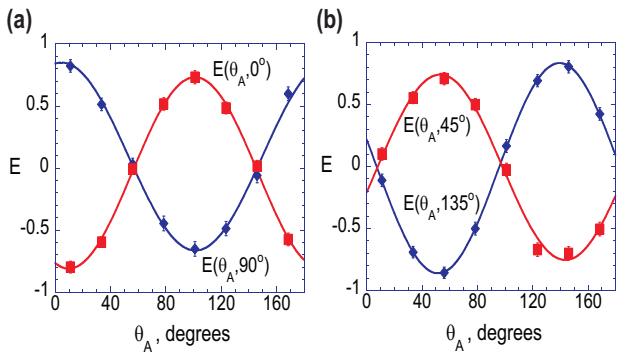,height=1.8in,width=3.2in}
\end{center}
\caption{Measured correlation function $E(\theta _A, \theta _B)$ as a function of $\theta _A$. (a),  $\theta _B=0^{\circ }$, squares, and
$90^{\circ }$, diamonds.  (b),  $\theta _B=45^{\circ }$, squares, and $135^{\circ }$, diamonds. The curves are sinusoidal fits to the data.
}\label{Fig2}
\end{figure}

Observation of Bell inequality violation is one method to confirm two-particle entanglement, by way of measurement of discrete values of $
C_{nm}\left(\theta _A,\theta _B\right) $ at polarization settings which lie on the slopes of the fringe pattern. Explicitly, following
Clauser-Horne-Shimony-Holt (CHSH) \cite{chsh}, we calculate the correlation function $E\left(\theta _A,\theta _B\right)$, given by
\begin{equation}
\frac{C_{13}\left(\theta _A,\theta _B\right)+C_{24}\left(\theta _A,\theta _B\right)-C_{14}\left(\theta _A,\theta _B\right)-C_{23}\left(\theta
_A,\theta _B\right)}{C_{13}\left(\theta _A,\theta _B\right)+C_{24}\left(\theta _A,\theta _B\right)+C_{14}\left(\theta _A,\theta
_B\right)+C_{23}\left(\theta _A,\theta _B\right)}.
\end{equation}

\begin{table}
\caption{\label{tab:table1} Measured values of the correlation function $E(\theta _A, \theta _B)$ at particular polarization settings and the
Bell parameter $S$.}
\begin{ruledtabular}
\begin{tabular}{ccccc}
$\theta _A$ & $\theta _B $& $E(\theta_A, \theta _B)$  \\
\hline
 78.5  & 45 &  $0.447 \pm 0.017$   \\
33.5 & 45     &  $0.640  \pm 0.014$   \\
78.5  & 0 &  $0.572  \pm 0.015$   \\
33.5  & 0    &  $-0.504  \pm 0.016$   \\
     &    & $S=2.16 \pm 0.03$   \\

\end{tabular}
\end{ruledtabular}
\end{table}
In Fig.~3 we display $E\left({\theta_A},\theta _B\right)$  as a function of $\theta _A$, for four values of $\theta _B$. By fitting the
correlation functions in Fig.~3 with sinusoids, we determine a set of four pairs of angles $\theta_A=78.5^{\circ }$, $\theta_B=45^{\circ }$,
$\theta_A^\prime=33.5^{\circ }$ and $\theta_B^\prime=0^{\circ }$ that should maximize the Bell inequality violation. We acquire data for two
hours at each of these four points (Table 1). In order to account for unequal efficiencies of the detectors $D1,D2$ and $D3,D4$, each correlation
measurement consisted of four runs, flipping polarization of either one of the idler fields by 90 degrees between the runs. As a result, the
products $\epsilon_m\epsilon _n$ are effectively replaced by the symmetric factor $\frac{1}{4}(\epsilon _1 + \epsilon _2)(\epsilon _3 + \epsilon
_4)$ in Eq. (5). In this case the correlation function $E\left(\theta _A,\theta _B\right)$ becomes independent of these efficiencies:
 \begin{eqnarray}
    E(\theta_A,\theta_B) &=& -\frac{1}{2} [ \cos(2(\theta_A-\theta_B))(1-\cos\phi_f
                           \sin 2\eta_f)  \nonumber \\
                         &+&  \cos(2(\theta_A+\theta_B))
               (1+\cos\phi_f \sin 2\eta_f) ].
  \end{eqnarray}

The CHSH version of the Bell inequality is then $|S|\leq 2$, where
\begin{equation}
S=E\left(\theta _A,\theta _B\right) + E\left(\theta_A^\prime,\theta _B\right)+E\left(\theta _A,\theta _B^\prime\right)-E\left(\theta
_A^\prime,\theta _B^\prime\right).
\end{equation}
We find $S=2.16 \pm 0.03 \nleq 2$, in clear violation of the Bell inequality. No corrections for background or dark counts were made to any of
the experimental counting rates, and these are chiefly responsible for the reduction in the observed value of $S$ from the ideal value of $2.60$
predicted by our theoretical model \cite{jenkins1}. In conclusion, we have demonstrated entanglement of two remote atomic qubits, based on
collective atomic states. By photoelectric detection of polarization correlations of the idler fields we have also confirmed the mapping of
atomic qubit entanglement onto photonic qubits. Long-lived entanglement of remote massive qubits and entanglement transfer between matter and
light are important prerequisites for realization of a scalable quantum information network.

This work was supported by Office of Naval Research, NASA, National Science Foundation, Research Corporation, Alfred P. Sloan Foundation, and
Cullen-Peck Chair. We thank M. S. Chapman for illuminating discussions and E.T. Neumann for experimental assistance.

\end{document}